\documentclass[10pt]{article}
\usepackage{graphicx}
\usepackage{wds11}

\lefthead{IVANYTSKYI ET AL.}
\righthead{Geometrical CLUSTERIZATION IN SU(2) GLUODYNAMICS}

\begin{document}

\title{Geometrical Clusterization in SU(2) gluodynamics and Liquid-gas Phase Transition}

\author{A. I. Ivanytskyi, K. A. Bugaev, V. V. Sagun, D. R. Oliinychenko}

\affil{Bogolyubov Institute for Theoretical Physics, Kyiv, Ukraine.}

\author{D. R. Oliinychenko}

\affil{FIAS, Frankfurt upon Main, Germany.}\vspace*{-2.mm}

\begin{abstract}
The liquid droplet formula is applied to an analysis of  the properties of   geometrical  (anti)clusters formed in SU(2) gluodynamics by the Polyakov 
loops of the same sign.  Using this  approach, we explain the phase transition in SU(2) gluodynamics as a transition between two liquids during 
which one of the liquid droplets (the largest cluster of a certain Polyakov loop sign) experiences a condensation, while another droplet (the next 
to the largest cluster of the opposite  sign of Polyakov loop) evaporates. The clusters  of smaller sizes form two accompanying gases, which behave oppositely to their liquids. The liquid droplet formula is used to analyze the size distributions of the gas (anti)clusters. The fit of these distributions 
allows us to extract  the temperature dependence of  surface tension and the value of  Fisher topological exponent $\tau$ for both kinds of gaseous 
clusters. It is shown that the  surface tension coefficient  of gaseous (anti)clusters can serve as an order parameter of the deconfinement phase 
transition in SU(2) gluodynamics. The Fisher topological exponent $\tau$ of (anti)clusters is found to have the same value $1.806\pm 0.008$. This 
value disagrees with the famous Fisher droplet model, but it agrees well with an exactly solvable model of  nuclear liquid-gas phase transition.  This 
finding may evidence for the fact that the SU(2) gluodynamics and this exactly solvable model of  nuclear liquid-gas phase transition are in the same 
universality class.\vspace*{-2.mm} \\ 

{\bf Kewords:} geometrical clusters, size distributions, liquid droplet  model formula, surface tension \vspace*{-5.mm}

\end{abstract}


\begin{article}

\section{Introduction}

The lattice formulation of quantum chromodynamics (QCD) is presently  considered  as the only first principle tool to investigate a transition between the  confined and  deconfined states of strongly interacting matter. Such a phase transition (PT) is also expected in a pure non abelian SU(N) gauge theory which is known as gluodynamics. The Svetitsky-Jaffe hypothesis [\markcite{{\it Yaffe and Svetitsky}, 1982}] relates the deconfinement PT in  (d+1)-dimensional SU(N) gluodynamics to the magnetic PT in the d-dimensional Z(N) symmetric spin model. The key element of this correspondence  is that the role of spin in the original SU(N) gluodynamics is played by the so called Polyakov loop. The latter is interpreted as the time propagator 
of an infinitely heavy static quark. A high level of understanding of the spin systems along with the Svetitsky-Jaffe hypothesis led  to 
a significant progress in studies of the SU(N) gluodynamics properties in the PT vicinity. Formation of geometrical clusters composed of the Polyakov loops is an important feature of the pure gauge theories [\markcite{{\it Fortunato and Satz}, 2000}, \markcite{{\it Fortunato et al.}, 2001}]. A similar phenomenon is well known in spin systems and it is responsible for percolation of clusters. Moreover, the deconfinement PT in gluodynamics already was 
studied within the percolation framework  in Refs. [\markcite{{\it Gattringer}, 2010}, \markcite{{\it Gattringer and Schmidt}, 2011}],  where the  main  attention  was paid to the largest and the next to the largest clusters whereas the  smaller clusters were ignored. However, in many respects the PT details are encoded in  the properties of smaller clusters which is well-known  after formulating  the  Fisher Droplet Model (FDM) [\markcite{{\it Fisher}, 1967,1969}]. 
An important finding  of the FDM is that at  the critical point the  size distribution of physical clusters obeys a 
power law which is controlled by the Fisher topological exponent $\tau$. 
Hence the value  of this exponent is rather important in order to develop a consistent
theory of PT in strongly interacting matter and to localize the  critical point of QCD  phase diagram which is  a hot topic of the  physics of heavy ion collisions. Therefore, in this work  we study 
the geometrical clusterization in SU(2)  gluodynamics and analyze the properties of clusters of all possible sizes. 
Such an  approach allows us to
explain the deconfinement of color charges as a specific kind of the liquid-gas PT [\markcite{{\it Ivanytskyi et al.}, 2016}]. 

\begin{figure}[t]
\begin{center}
\begin{tabular}{c}
\includegraphics[width=78mm]{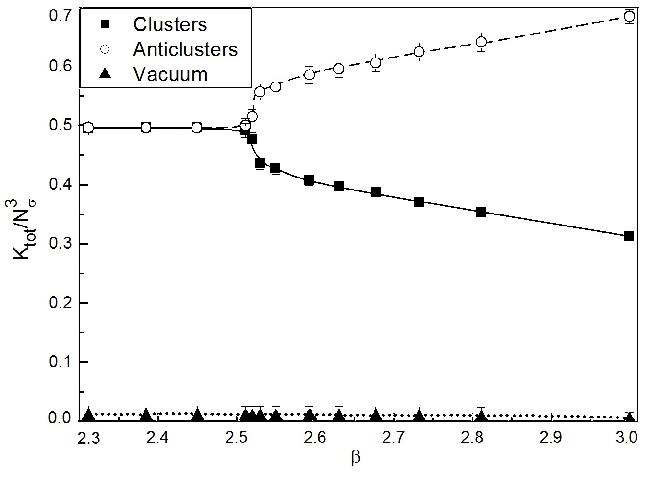}
\includegraphics[width=7.8cm]{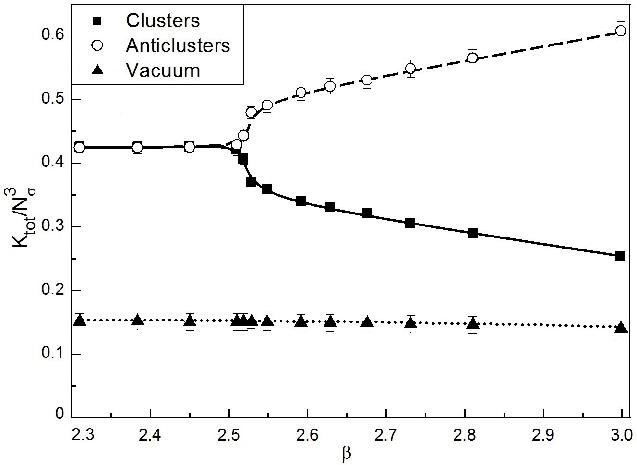}
\end{tabular}
\end{center}\vspace*{-6.mm}
\caption{Total volume fraction of clusters, anticlusters and auxiliary vacuum measured for $L_{cut}=0.1$ (left panel) and $L_{cut}=0.2$ (right panel).
The volume fraction of an  auxiliary vacuum is independent on the Polyakov loop cut off. The curves are shown to guide the eye.}\vspace*{-3.mm}
\end{figure} 

\section{The Polyakov loop geometrical clusters}

The local Polyakov loop is  a gauge invariant analog of continuous spin which exists in every   space point of the lattice. In  case of discrete (d+1)-dimensional
lattice of size $ N_\sigma^d\times N_\tau$  it is defined by the trace of temporal gauge links $U_4(\vec x, t)$
\begin{equation}
L(\vec x)=Tr\prod_{t=0}^{N_\tau-1} U_4(\vec x, t).
\end{equation}
In case of $SU(2)$ gauge group the quantity $L(\vec x)$ has real values  in the interval from  $-1$ to $1$. For a given lattice configuration 
the neighboring  Polyakov loops can be attributed to one cluster,  if they have the same sign. The boundaries of clusters with opposite signs 
of  Polyakov loop are characterized by the strong fluctuations of $L(\vec x)$. Therefore, similarly to Refs. [\markcite{{\it Gattringer}, 2010}, 
\markcite{{\it Gattringer and Schmidt}, 2011}] we introduced the minimal absolute value of the Polyakov loop attributed to the clusters, i.e. a 
cut off parameter $L_{cut} > 0$. All space points $\vec x$ with $|L(\vec x)| \le L_{cut}$ are attributed  to ``auxiliary'' or ``confining'' vacuum 
which volume fraction is independent of the inverse lattice coupling $\beta=\frac{4}{g^2}$  (see Fig. 1), where $g^2$ is the lattice coupling 
constant. The above definition allows us to  define the monomers, the dimers, etc. as the clusters made of a corresponding number of ``gauge 
spins" of the same sing which are  surrounded  either by the  clusters of the opposite ``spin''  sign or  by a vacuum [\markcite{{\it Ivanytskyi et al.}, 2016}].  Obviously, there are  clusters  of two types related to two signs  of the local Polyakov loop. We introduce a formal definition of  anticlusters, if  
their sign coincides with the sign of the largest n-mer existing at a given lattice configuration. The largest anticluster is called the ``anticluster 
droplet" whereas the  other n-mers of the same sign correspond to the ``gas of anticlusters". The  clusters are defined to  have an  opposite 
sign of the Polyakov loop,  whereas the largest of them is called  the ``cluster droplet".

\begin{figure}[t]
\begin{center}
\begin{tabular}{c}
\includegraphics[width=15.6cm]{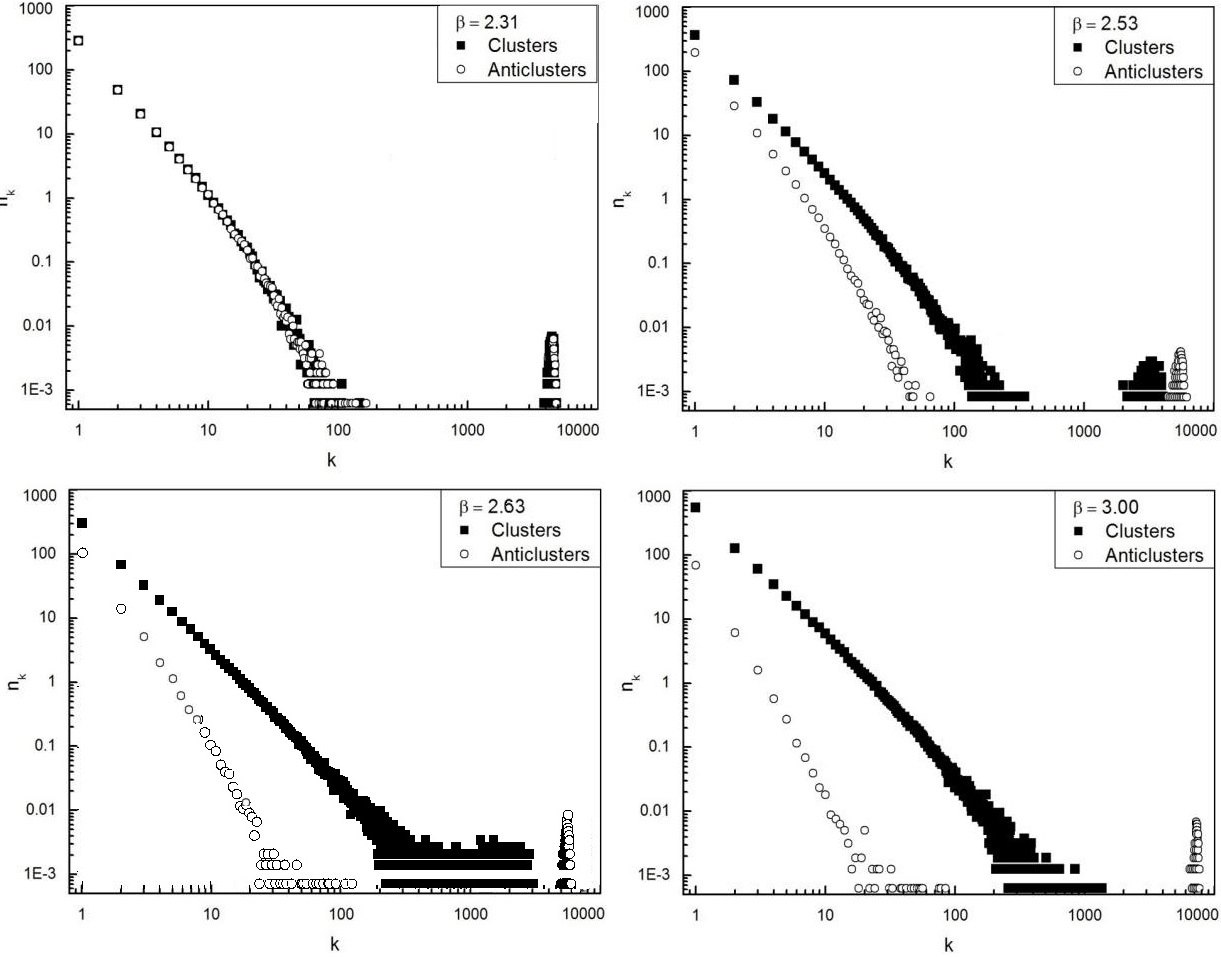}
\end{tabular}
\end{center}\vspace*{-6.mm}
\caption{Size distributions of clusters (squares) and anticlusters (circles) for $\beta=2.31$ (left upper panel), $\beta=2.53$ (right upper panel),
$\beta=3.0$ (lower panel) are shown for the  cut off $L_{cut}=0.2$. }\vspace*{-3.mm}
\end{figure}

\subsection{Size distributions of clusters}

The described  scheme  of the (anti)cluster identification was realized numerically. The Polyakov loops were obtained at each spatial point of 3+1 dimensional lattice with the spatial and temporal extents   $N_\sigma=24$ and $N_\tau=8$, respectively. The simulations were performed for 13 values of the inverse lattice coupling $\beta$ inside the interval $\beta\in[2.31,3]$. The physical temperature
$T$ is unambiguously defined by the temporal extent and the lattice spacing $a$ as $1/T=N_\tau a$.  The $\beta$ points where distributed not uniformly. They where concentrated in the PT region which is of principal interest for  this study. The identification of (anti) clusters was performed for two values of the Polyakov loop cut off $L_{cut}=0.1$ and $0.2$. For almost all $\beta$ points the number of (anti)clusters of each  size was averaged over  the ensemble of 800 and 1600 gauge field lattice configurations. 
The  distributions  obtained in this way were  practically the same  (within the statistical errors).
Hence, in our analysis we  used the results found  for 1600 configurations as the high statistics limit. 
In case of gaseous anticlusters at three largest values of  $\beta$ the  statistics was rather poor. Therefore, for these points we used   2400 configurations. The right hand side vicinity of  PT  was also analyzed with higher  statistics in order to  exclude the  effects of statistical fluctuations which we observed for $\beta=2.52,~2.53$ and $2.67$. 
The typical size distributions of (anti)clusters $n_k$ for $L_{cut}=0.2$  are shown in Fig. 2.
For the lattice constant values  below the critical value  $\beta_c^\infty=2.5115$  in an  infinite system 
[\markcite{{\it Fingberg et al.}, 1993}]  the distributions of (anti)clusters are identical due to existing  global Z(2) symmetry (the left upper panel of Fig. 2). 
If $\beta$ is even slightly above $\beta_c^\infty$ then the symmetry between (anti)clusters breaks down (the right upper panel of Fig. 2). In this case the  size  of the cluster (anticluster) droplet decreases (increases). It is remarkable, that the corresponding gaseous clusters  behave contrary to their droplets. Therefore,  the deconfinement PT in SU(2) gluodynamics  can be considered as an  evaporation of the cluster droplet into the gas of clusters  and  a simultaneous condensation of the gas of  anticlusters  into the anticluster 
droplet. At high $\beta$ (the lower panel of Fig. 2) the cluster droplet becomes indistinguishable from 
its gas whereas the size of  anticluster droplet becomes comparable to the size of  system.  From Fig. 2  one can see that the gas and liquid  branches of anticluster distributions are well separated from each other. A
similar behavior is seen for the clusters, except for  very  high values of  $\beta$,  where  the cluster droplet simply disappears due to evaporation (see the lower panel of Fig. 2).

\begin{figure}[t]
\begin{center}
\begin{tabular}{c}
\includegraphics[width=16cm]{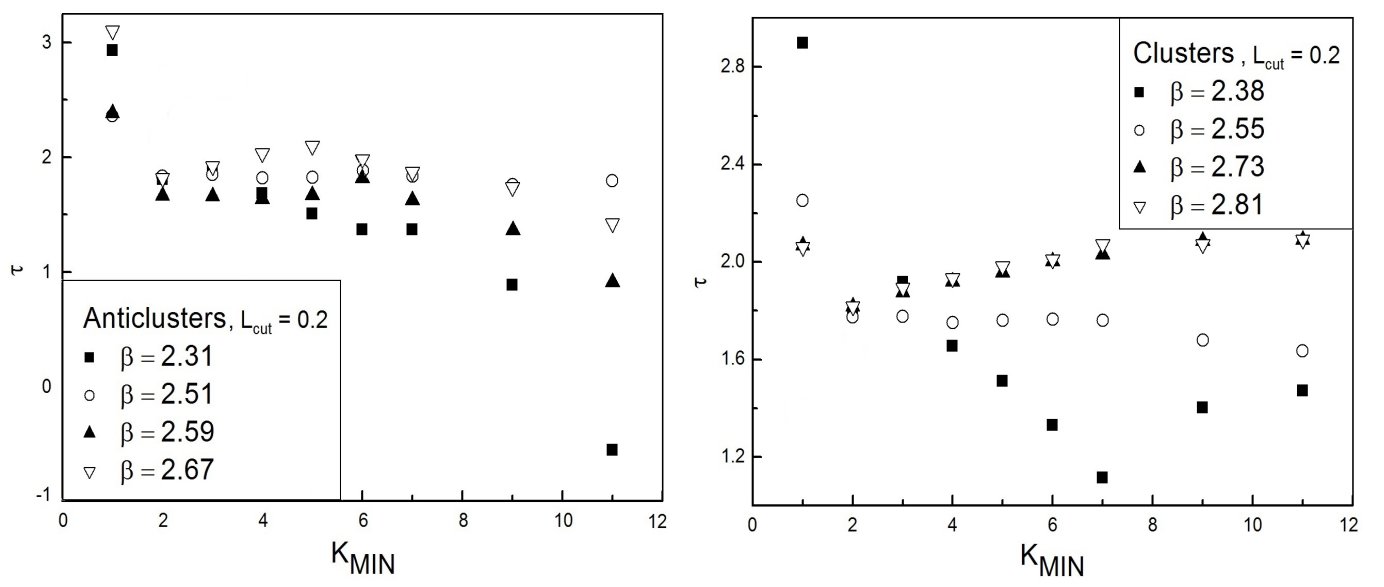}
\end{tabular}
\end{center}\vspace*{-6.mm}
\caption{The Fisher exponent $\tau$ for several values of $k_{min}$ and for a few values of $\beta$ found by the 4-parametric fit 
of the LDM formula. All these results refer to the cut-off $L_{cut}=0.2$. }\vspace*{-3.mm}
\end{figure}

Since the found distributions closely resemble the ones discussed  for the nuclear fragments   and for the Ising  spin clusters  
[\markcite{{\it Sagun et al.}, 2014}], we  decided to study the question  whether 
the Liquid Droplet Formula
(LDF) [\markcite{{\it Fisher}, 1967}] 
\begin{equation}
\label{EqII}
n_A^{th}(k)=C_A\exp\left(\mu_A k-\sigma_A k^{2/3}-\tau \ln k\right) \,,
\end{equation}
is able to reproduce the size distributions  of clusters ($A=cl$) and anticlusters ($A=acl$). Here $\mu_A k$ is the bulk part of free energy of   k-mer (anti)cluster, $\sigma_A k^{2/3}$ denotes its surface free energy with the surface proportional to $k^{2/3}$, $\tau$ is the Fisher topological constant and $C_A$ is the normalization factor. These parameters were defined from the fit of 
the (anti)cluster size distributions.  Since the LDM is valid only for (anti)clusters which are  sufficiently large, then the question which has  to be clarified  first  was a determination of the minimal size $k_{min}$ to  which Eq. (\ref{EqII}) can be applied. A minimization of the quantity 
\begin{equation}
\label{EqIII}
\frac{\chi^2}{dof}=\frac{1}{k_{max}-k_{min}-4}\sum_{k=k_{min}}^{k_{max}}\left(\frac{n_A^{th}(k)-n_A(k)}{\delta n_A(k)}\right)^2 \,,
\end{equation}
with respect to $C_A$, $\mu_A$, $\sigma_A$, $\tau$ and $k_{min}$ allowed us to answer this question. Here $k_{max}$ is the maximal size of 
the gaseous (anti)cluster  and $\delta n_A(k)$ denotes the error of $n_A(k)$ obtained in simulations. In  practice, the minimization procedure was 
performed using the iterative gradient search method,  in which  the next approximation of the parameter vector $\vec p_A=(C,\mu,\sigma,\tau)_A$ 
is defined as
\begin{equation}
\vec p_A\rightarrow\vec p_A-\epsilon\cdot\vec\nabla_{\vec p_A}\frac{\chi^2}{dof} \;,
\end{equation}
while the  positive valued elements of the matrix $\epsilon=diag(\epsilon_C,\epsilon_\mu,\epsilon_\sigma,\epsilon_\tau)$ are used  in order to optimize
the fit procedure. Evidently, this iterative scheme corresponds to a numerical solution of  the equation $\vec\nabla_{\vec p_A}\frac{\chi^2}{dof}=0$ 
which is a criterion of  $\frac{\chi^2}{dof}$ minimization. A comparison of the fit quality obtained for different $k_{min}$ values  shows  that the LDF is
already  able to describe the dimers. Indeed, we found that $\frac{\chi^2}{dof}\approx 1$  for all $k_{min}\ge2$ , whereas for $k_{min}=1$  we got 
$\frac{\chi^2}{dof} \approx 10$ which corresponds to the low quality of  data description. Our analysis of the Fisher topological constant behavior as 
a function of $\beta$ demonstrates  that for $k_{min}=2$ it is, indeed, a  constant within the statistical errors (see Fig. 3). This remarkable result is in 
line with predictions of the cluster type models [\markcite{{\it Fisher}, 1967,1969}, \markcite{{\it Sagun et al.}, 2014}]. Moreover,  we found  that 
$\tau<2$ both  for clusters and for  anticlusters. This result agrees  with the  exactly solvable model of the nuclear liquid-gas phase transition 
[\markcite{{\it Sagun et al.}, 2014}] and contradicts to the FDM [\markcite{{\it Fisher}, 1967,1969}]. Thus, performing a four parametric fit of the LDF 
we found that $k_{min}=2$ and $\tau=1.806\pm0.008$ both for  clusters and for anticlusters. After fixing the value of  $\tau$ exponent, we used 
a three parametric fit of the (anti)cluster size distributions  to define $C_A$, $\mu_A$ and $\sigma_A$ with high precision. This was done by minimizing   
$\frac{\chi^2}{dof}$ for   $k_{min}=2$ and $\tau=1.806$.  The typical value of $\frac{\chi^2}{dof} \approx 1$ was obtained for any  $\beta$, which
signals  about the high quality of the data description. The $\beta$-dependences of $\mu_A$ and $\sigma_A$ are shown in Fig. 4 for $L_{cut}=0.2$. 
For $L_{cut}=0.1$ the results are qualitatively the same. 

\begin{figure}[t]
\begin{center}
\begin{tabular}{c}
\includegraphics[width=16cm]{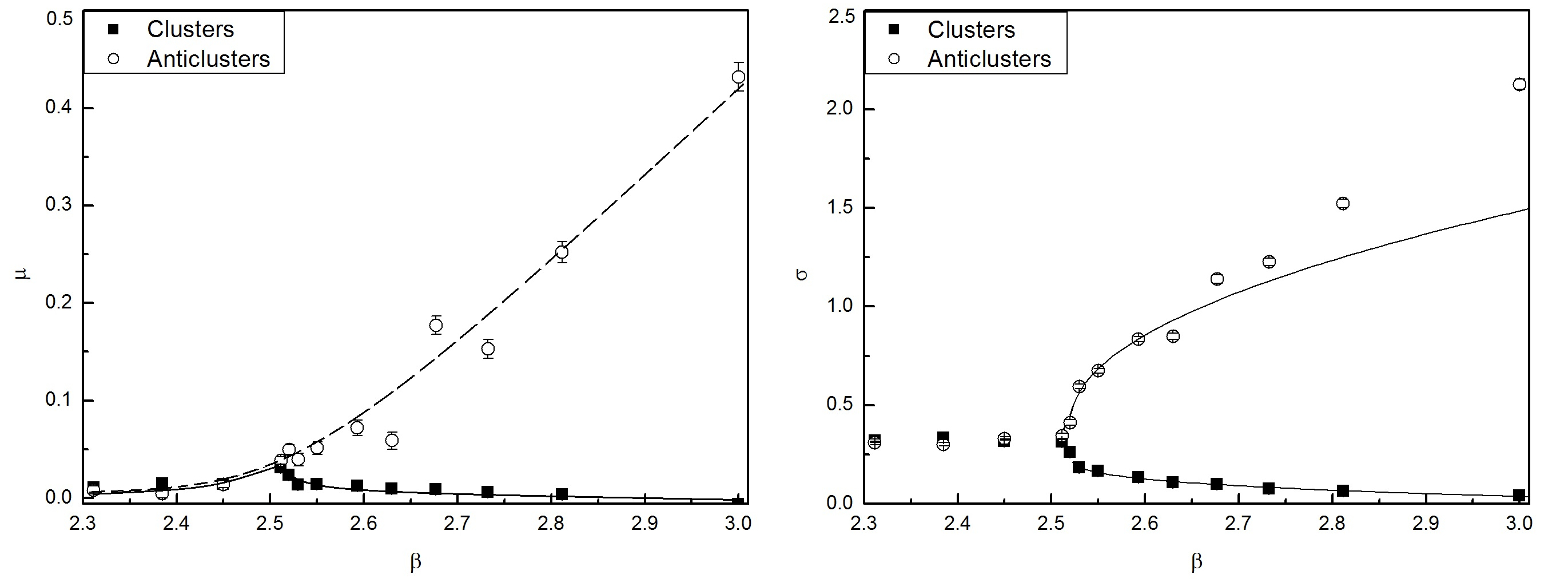}
\end{tabular}
\end{center}\vspace*{-6.mm}
\caption{Dependence of the reduced chemical potential $\mu_A$ (left panel) and the reduced surface tension $\sigma_A$ (right panel) on the 
inverse lattice coupling $\beta$ obtained for $L_{cut}=0.2$. The curves shown  in the right panel represent Eq. (\ref{V}).}\vspace*{-3.mm}
\end{figure}

\subsection{Order parameters of the deconfinement  phase transition}

From the right panel of Fig. 4 one can see  that the behavior of the reduced surface tension coefficient $\sigma_A$ drastically changes when $\beta$ exceeds the critical value
$\beta_c=2.52$. Indeed, for  $\beta\le\beta_c$ this quantity is constant and it is identical for clusters and anticlusters, 
while for $\beta>\beta_c$ the situation 
is completely different,  since $\sigma_{cl}$  monotonically decreases and $\sigma_{acl}$ monotonically  increases with $\beta$. The qualitatively 
different  behavior of these quantities allows us to treat them as an order parameter of the deconfinement PT in SU(2) gluodynamics. The $\beta$-dependence 
of the reduced surface tension in the right  hand side vicinity of $\beta_c$ can be successfully  described by the exponent $B_A$ as
\begin{equation}
\label{V}
\sigma_A(\beta)=\sigma_A(\beta_c)\pm d_A (\beta-\beta_c)^{B_A},
\end{equation}
where the signs ``+" and ``-" correspond to A=acl and A=cl, respectively,   
and $d_A$ is the normalization constant. The
$\beta$-dependence  of $\sigma_A$ was described by this formula. The results of fit are given in Table 1. Despite the different values of  $d_A$ found for different cut offs, the exponents $B_A$ do not show any dependence on $L_{cut}$.

\begin{table}[htdp]\vspace*{-9.mm}
\caption{The fit parameters according to Eq. (\ref{V}).}
\begin{center}
\begin{tabular}{|c|c|c|c|c|}
\hline
~ {\rm  cut-off}~  &  ~ {\rm  type}~   &  $d_A$     &    $B_A$   &   $\chi^2 / dof$    \\
\hline
  ~~ $L_{cut} = 0.1 $ ~~ &  ~ {\rm clusters} ~ &~~$0.485 \pm 0.014$~~ &   ~~$0.2920 \pm 0.0012$~~  &  ~~$1.43 /4 \simeq 0.36$~~ \\ \hline 
$L_{cut} = 0.1 $   &    ~ {\rm anticlusters} ~ &  $2.059 \pm 0.028$  &   $0.4129\pm 0.0077$  & ~~$1.68/4\simeq 0.48 $~~ \\ \hline
$L_{cut} = 0.2 $ &     ~ {\rm clusters} ~ & $0.2796 \pm 0.0118$   &   $0.2891 \pm 0.0016$ &  $1.11/4 \simeq  0.28$ \\ \hline 
$L_{cut} = 0.2 $&     ~ {\rm anticlusters} ~ & $1.344 \pm 0.033$  &    $0.4483 \pm 0.0021$  & $0.66/2 \simeq 0.33$   \\ \hline
\end{tabular}
\end{center}
\label{table3}
\end{table}
\vspace*{.3cm}
The mean size of the droplet we found according to the expression
\begin{equation}
\label{EqVI}
\max K_A=\frac{\sum\limits_{k=1} k^{1+\tau}n_A(k)}{\sum\limits_{k=1} k^{\tau}n_A(k)} \,. 
\end{equation}
The $\beta$-dependence of  $\max K_A$  is shown in Fig. 5. Its behavior for $\beta>\beta_c$
is described by the  exponent $b_A$
\begin{equation}
\label{VII}
\max K_A(\beta)=\max K_A(\beta_c)\pm a_A (\beta-\beta_c)^{b_A},
\end{equation}
where the signs ``+" and ``-" correspond to A=acl and A=cl, respectively,  and $a_A$ is the normalization constant. The parameters  found by the fit are given in  Table 2.

\begin{table}[htdp]\vspace*{-9.mm}
\caption{The fit parameters according to Eq. (\ref{VII}).}
\begin{center}
\begin{tabular}{|c|c|c|c|c|}
\hline
~ {\rm  cut-off}~  &  ~ {\rm  type}~   &  $a_A$     &    $b_A$   &   $\chi^2 / dof$    \\
\hline
  ~~ $L_{cut} = 0.1 $ ~~ &  ~ {\rm clusters} ~ &~~$3056 \pm 246$~~ &   ~~$0.2964 \pm 0.0284$~~  &  ~~$16.32/4 \simeq 4.08$~~ \\ \hline 
$L_{cut} = 0.1 $   &    ~ {\rm anticlusters} ~ &  $2129 \pm 160$  &   $0.3315 \pm 0.0269$  & ~~$8.94/4 \simeq 2.235$~~ \\ \hline 
$L_{cut} = 0.2 $ &     ~ {\rm clusters} ~ & $4953 \pm 443$  &   $0.3359 \pm 0.0289$ &  $12.3/3 \simeq 4.01$ \\ \hline 
$L_{cut} = 0.2 $&     ~ {\rm anticlusters} ~ & $2462 \pm 87.7$  &    $0.3750 \pm 0.0129$  & $2.068/4 \simeq 0.517$   \\ \hline 
\end{tabular}
\end{center}
\label{table1}
\end{table} \vspace*{.3cm}

It is remarkable that the found exponents $b_A$ are close to the critical exponent $\beta_{Ising}=0.3265\pm0.0001$ of the 3-dimensional Ising model
 [\markcite{{\it Campostrini et al.}, 2002}] and to the critical exponent $\beta_{liquids}=0.335\pm0.015$ of simple liquids  
[\markcite{{\it Huang}, 1987}],  whereas for $L_{cut}=0.2$ the exponent $b_{acl}$ is somewhat larger. 

\begin{figure}[t]
\begin{center}\vspace*{-1.mm}
\begin{tabular}{c}
\includegraphics[width=7.8cm]{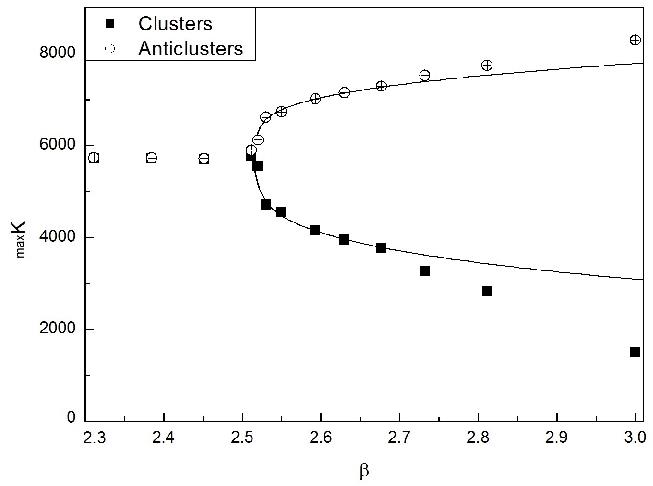}
\includegraphics[width=7.8cm]{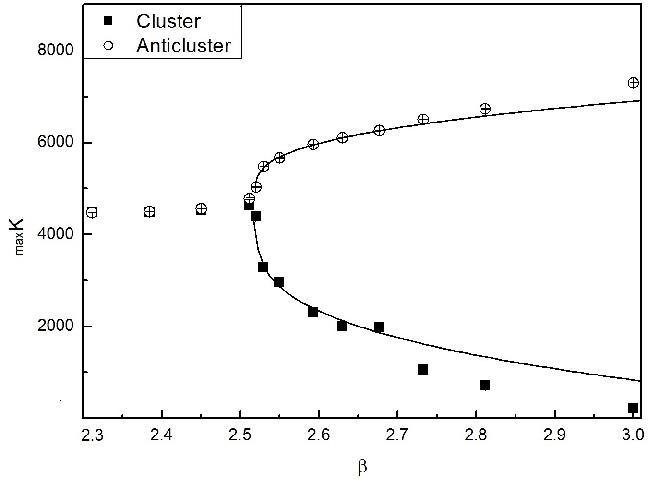}
\end{tabular}
\end{center}\vspace*{-8.mm}
\caption{Dependence of the mean  size of the maximal (anti)cluster found for $L_{cut}=0.1$ (left panel) and for $L_{cut}=0.2$ (right panel). 
The curves represent Eq. (\ref{VII}).}\vspace*{-4.mm}
\end{figure}

\section{Conclusions}

In this contribution  we present a novel approach to study the deconfinement PT  in the SU(2) pure gauge theory in terms of the geometrical
clusters composed of the Polyakov loops of the same sign. We demonstrate  that  the separation of (anti)clusters into ``liquid" droplet and 
``gas" of smaller fragments is well justified and reflects the physical properties of the lattice system. This concept allows  us to explain the 
deconfinement PT as a special kind of the liquid-gas transition. However, in contrast to the ordinary liquids,  the SU(2) gluodynamics contains  
two types of liquid whose behavior is drastically  different in the region of broken global Z(2) symmetry. The cluster liquid droplet evaporates 
above  PT whereas the anticluster liquid droplet experiences the condensation of the accompanying gas of anticlusters. A successful application 
of the LDF to the description of the size distributions of gaseous (anti)clusters  is the main  result of this study. Surprisingly, even the monomers 
are qualitatively  described by Eq. (\ref{EqII}). The fit  of  the (anti)cluster size distributions by the  LDF formula allows us to determine  the 
$\beta$-dependences  of the reduced chemical potential and the reduced surface tension coefficient. While in the symmetric phase this quantities 
are identical for fragments of  both kinds,  their behavior is drastically different in the deconfined phase. Another  important finding of this study 
is a high precision determination of the Fisher topological constant $\tau=1.806\pm0.008$ which is the same  both for  clusters and for anticlusters.
This result is in line with the exactly solvable model of the nuclear liquid-gas PT [\markcite{{\it Sagun et al.}, 2014}]. At the same time it  disproves the 
FDM prediction  that $\tau > 2$ [\markcite{{\it Fisher}, 1968}]. We showed  that  the reduced surface tension coefficient and the  mean  size of the 
largest (anti)cluster can be used as  the new order parameters of  deconfinement  PT in SU(2) gluodynamics. In contrast to the FDM the power law 
for the size distribution  is found only for the gas of clusters and not at the PT point,  but at $\beta=3$. \vspace*{-2.mm}

\vspace*{-.1cm}

\acknowledgments 
{The authors thank D. B. Blaschke, O. A. Borisenko, V. Chelnokov, Ch. Gattringer, D.
H. Rischke, L. M. Satarov, H. Satz and E. Shuryak for the fruitful discussions and
valuable comments. The present work was supported in part by the National Academy of
Sciences of Ukraine and by the NAS of Ukraine grant of GRID simulations for high
energy physics.} \vspace*{-2.mm}



\end{article}

\end{document}